# Phase diagram of microcavity polariton condensates with a harmonic potential trap

Ting-Wei Chen[1], Ming-Dar Wei[1], Szu-Cheng Cheng[2,*], Wen-Feng Hsieh[1,3,†]

[1]Department of Photonics, National Cheng Kung University, Tainan, Taiwan
[2]Department of Optoelectric Physics, Chinese Culture University, Taipei, Taiwan
[3]Department of Photonics and Institute of Electro-Optical Engineering, National Chiao Tung University, Hsinchu, Taiwan
*sccheng@faculty.pccu.edu.tw; †wfshieh@mail.nctu.edu.tw

Abstract

We theoretically explore the phase transition in inhomogeneous exciton-polariton condensates with variable pumping conditions. Through Bogoliubov excitations to the radial-symmetric solutions of complex Gross-Pitaevskii equation, we determine not only the bifurcation of stable and unstable modes by the sign of fluid compressibility but also two distinct stable modes which are characterized by the elementary excitations and the stability of singly quantized vortex. One state is the quasi-condensate BKT phase with Goldstone flat dispersion; the other state is the localized-BEC phase which exhibits linear-type dispersion and has an excitation energy gap at zero momentum.

# I. Introduction

It's known that there is no Bose-Einstein condensate (BEC) in the thermodynamic limit at any finite temperature in homogeneous two-dimensional (2D) systems, and instead superfluidity can be achieved through Berezinskii-Kosterlitz-Thouless (BKT) transition in an interacting system. The emergence of superfluidity was interpreted by the formation of a topological order resulting from vortex-antivortex pairs (VAPs) [1-3]. The BKT theory for a uniform system predicts a density jump of superfluid at the transition [4], namely, the transition happens at a universal ratio of $n_s \lambda_T^2 = 4$, where $n_s$ is the superfluid density and $\lambda_T$ is the thermal de Broglie wavelength of BEC. However, the universality for 2D trapped systems was experimentally observed to be 6 ± 2 in cold atoms [5].

In parallel to the pursuing of BEC in semiconductor microcavities [6-9], microcavity polariton condensate (MPC) was proposed as a candidate to study superfluid, which is particularly characterized by the formation of vortices [10-12], collective dynamics as well as elementary excitations with linear dispersion [13]. In analogy to the trapped 2D cold atoms, the major difference of MPCs is the local self-equilibrium property coming from the finite lifetime of exciton polarions. Inhomogeneous 2D polariton BECs with a continuous replenishment have been theoretically studied in use of the modified complex Gross-Pitaevskii equations (cGPEs) [14]. The thermodynamic properties of 2D exciton-polaritons had been analyzed and shown to exhibit local condensation or BKT phase transition towards superfluidity [15]. In addition, the dynamical evolution of the condensate was investigated by a quantum kinetic formalism with the distribution function of polaritons described by a semi-classical Boltzmann equation and phase diagrams for several microcavities were obtained with respect to the temperature and polariton density [15]. Similar phase diagram had been calculated with randomly distributed disorders based on cGPEs, in which quasi-particle excitation spectra were also discussed for different phases [16]. Furthermore, the BKT-like phase was investigated that used spatial correlation to observe the power-law decay indicating the existence of BKT-like phase in open-dissipative systems [17].

Instead of considering the random disorders, we show in this paper by studying quasi-particle excitations to the cGPE with a harmonic trap assuming the relatively slow dynamics of the condensate by eliminating the reservoir effect so that the dynamics can be easily studied by a single partial differential equation. In use of the cGPE accompanied with Bogoliubov excitations for studying the 2D polariton condensates in a harmonic trap, a bifurcation of stable and unstable modes can be obtained associated with the sign of fluid compressibility. Apart from the criterion for instability of the vortex-free state, we conclude a 2D trapped polariton condensate may exhibit both the BEC and BKT-like phases, separated by what we believe to be a quasi-condensation boundary. The phase diagram is established through the pumping spot and strength rather than the temperature and polariton density in those articles aforementioned.

## 2. Condensate wave functions and elementary excitations

The GPE is a mean-field model to deal with the many-body problems [18-19]. The classical field $\psi$, known as the macroscopic wavefunction of the polariton condensate, is used to replace the field operator by ignoring the non-condensate part from quantum and thermal fluctuations. Under the assumption that the polariton gas is weakly-interacting, the condensate dynamics can be described by the cGPE without microscopic physics of the polaritons involved. Here, a theoretical model is presented for a 2D finite system of MPCs as

$$i\hbar\frac{\partial \psi}{\partial t} = (-\frac{\hbar^2}{2m}\nabla^2 + V_{ext} + U|\psi|^2)\psi + i(\gamma_{eff} - \Gamma|\psi|^2)\psi, \tag{1}$$

where the external trapping potential is given by the harmonic form of $V_{ext}=m\omega^2 r^2/2$, with $\omega$ being the angular frequency, m the polariton mass, $r$ the radial coordinate. $U$ is the interaction energy similar to the optical Kerr constant; $\gamma_{eff}$ is the linear net gain describing the balance of the stimulated scattering of polaritons into the condensate and the decay of polaritons out of the cavity. This is actually a spatially dependent gain determined by the pumping scheme and $\Gamma$ represents the coefficient of nonlinear scattering loss. We assume the pump profile has a form of polar symmetric step-function with amplitude $\gamma_{eff}$ and pump spot size $R$.

Let the oscillator length be $L = \sqrt{\hbar/m\omega}$, so that the length, time and energy are in units of $L$, $1/\omega$ and $\hbar\omega$, respectively. Under Thomos-Fermi scaling, we rescale $\psi_k$ to be $\sqrt{\hbar\omega/2U}\Phi$ and introduce the dimensionless variables $\alpha = 2\gamma_{eff}/\hbar\omega$ and $\sigma = \Gamma/U$ to represent the pumping strength and scattering loss. The mean-field cGPE can be written as

$$i\frac{\partial \Phi}{\partial t} = [-\frac{1}{2}\nabla_{\vec{\rho}}^2 + \frac{\rho^2}{2} + \frac{1}{2}|\Phi|^2 + \frac{i}{2}(\alpha - \sigma|\Phi|^2)]\Phi . \qquad (2)$$

Here the Laplacian operator $\nabla_{\vec{\rho}}^2$ is associated with the dimensionless polar coordinate $\vec{\rho} = (\rho, \theta)$ with $\rho = |\vec{r}|/L$. With σ fixed throughout this paper, the nature of the system can be quantified by two control parameters, i.e., $\alpha$ and $R$, which are both accessible experimentally. The chemical potential $\mu(T)$ at quasi-thermal equilibrium temperature $T$ is related to the radially symmetric wavefunction by $\Phi(\rho,t) = \Phi(\rho)e^{-i\mu(T)t}$, and it will be determined once the pumping condition is specified. Therefore, the cGPE can be written as

$$[-\frac{1}{2}\nabla_{\vec{\rho}}^2 + \frac{\rho^2}{2} + \frac{1}{2}|\Phi|^2 + \frac{i}{2}(\alpha - \sigma|\Phi|^2)]\Phi = \mu(T)\Phi . \qquad (3)$$

The steady-state nonlinear GPE is numerically solved in use of the Runge-Kutta evolution and the shooting method with proper boundary conditions and additional constraint to enforce the population conservation of pumping and loss. The chemical potential for a given pumping scheme is targeted to fulfill both the boundary conditions and the conservation law of polaritons $\int(\alpha - \sigma\Phi^2)|\Phi|^2 d^2r = 0$, i.e., the balance between net gain and loss over all the space should be zero.

Once the stationary solutions are obtained, the stability can be investigated by means of Bogoliubov-de Genned analysis. We consider a small deviation $\delta\Phi = e^{-i\mu t}[w(\rho,\theta)e^{-i\Omega t} - v^*(\rho,\theta)e^{i\Omega t}]$ from the radially symmetric stationary solution $\Phi_0(\rho) = A(\rho)e^{i\varphi(\rho)}e^{-i\mu t}$, where $w(\rho,\theta) = W(\rho,\theta)e^{i\vec{k}\vec{\rho}}e^{i\varphi(\rho)}$ and $v(\rho,\theta) = V(\rho,\theta)e^{i\vec{k}\vec{\rho}}e^{i\varphi(\rho)}$; then we substitute the total wavefunction $\Phi = \Phi_0 + \delta\Phi$ into Eq. (3). The response of the polariton condensate under a small perturbation is calculated by linearizing the cGPE around the stationary solution in the momentum space. We assume the fluctuation is of an angular hormonic function multiplied by a radial Bessel function such that $W(\rho,\theta) = \sum_{m,n}\frac{A_{mn}}{2\pi}e^{im\theta}J_m(k_{mn}\rho)$ and $V(\rho,\theta) = \sum_{m,n}\frac{B_{mn}}{2\pi}e^{im\theta}J_m(k_{mn}\rho)$ where $k_{mn}$ multiplied by the size of the system $R_0$, i.e., $k_{mn}R_0$, represents the nth zero root of the mth-order Bessel function $J_m(\rho)$. The Bogoliubov equations derived this way couple quasi-particle amplitudes of $w_k$ and $v_k$ with the wavevector $k$ at excitation frequency $\Omega$, then the amplitudes $W$ and $V$ satisfy the following coupled equations:

$$L[W] - \frac{1}{2}(A^2 - i\sigma A^2)V = (\mu + \Omega)W, \quad (4)$$

$$L^+[V] - \frac{1}{2}(A^2 + i\sigma A^2)W = (\mu - \Omega)V, \quad (5)$$

where

$$L = -\frac{1}{2}e^{-i\varphi}\frac{1}{\rho}\frac{d}{d\rho}(\rho\frac{d}{d\rho}e^{i\varphi}) + [\frac{k^2}{2} + \frac{\rho^2}{2} + A^2 + \frac{i}{2}\alpha - i\sigma A^2] \quad (6)$$

and its adjoint operator $L^+$ are linear operators acting on $W(\rho)$ and $V(\rho)$, respectively. The excitation frequency $\Omega$ is a complex parameter with its real part standing for the oscillation frequency and imaginary part for the damping rate. The collective excitations are therefore the solutions of the eigenvalue problem described by Eq. (4) and Eq. (5), and we can express the Bogoliubov excitation spectra in terms of the linear momentum $k$ and the corresponding eigenfrequency $\Omega$.

We justify the stability of the eigenstate by seeking the eigenvalues of the coupled Bogoliubov equations. In this framework, the eigenstate with positive imaginary part is unstable; we call these unstable modes as the "dynamical instability modes". Figure 1(a) is the excitation spectrum of an instability mode derived from the pumping condition of $R = 6\,L$ and $\alpha = 4.4$. It shows the roton-maxon-like behavior with a turning point around $k = 1$. For a lower pumping strength, the excitation spectra follow a modified Bogoliubov dispersion shape, namely, of phonon-like linear dispersion in the low-momentum regime ($|k| < 1$), and quadratic in the free-particle regime ($|k| > 1$) as shown in Fig. 1(b). Furthermore, there exist two different stable modes with non-vanishing negative imaginary part, Im($\Omega$) < 0, for all linear momenta $k$'s. In Fig. 1(c), the stable modes exist for pumping schemes with zero real-part eigenvalue at zero momentum, i.e., Re[$\Omega(k = 0)$] = 0 or the zero-momentum excitation energy (ZMEE) equals to zero. In addition, it shows a Goldstone flatness of real-part eigenvalue for small k which is not expected from the Bogoliubov dispersion law and is a unique phenomenon for the non-equilibrium phase transitions [20, 21]. We call this kind of stable modes as the "soft" stable modes because it is relatively easier to be excited than the "rigid" stable modes. The spectrum of single particle excitation in a non-equilibrium polariton condensate has been theoretically studied for incoherent pumping [20, 22]. On the basis of the Keldysh Green's function [22] or the generalized Gross-Pitaevskii equation in the spatially homogeneous case [20], the collective modes are altered in the presence of pumping and decay that shows the low energy Goldstone mode. Here the Goldstone mode still exists in the spatially inhomogeneous case, and the diffusive behaviour at small $k$ indicates the soft mode is not superfulid. In Fig. 1(d), the rigid stable modes exist with nonzero ZMEE, i.e., Re[$\Omega(k = 0)$] $\neq$ 0, which implies a finite amount of energy is required for the excitation, and the corresponding excitation is massive compared to the soft stable mode. More obvious linear-type dispersion can be observed in the rigid mode depending on the pump conditions.

## 3. Phase diagram of the cavity polaritons in a harmonic potential trap

Figure 2 is a phase diagram as functions of pump spot and strength. At large pump spots and strengths, the boundary of stable-to-instable transition starts to deviate from the Thomas-Fermi (TF) approximation because the flux of the condensate could affect the density distribution, and the density gradient in the cGPE cannot be negligible anymore. It's also interesting to note that the stable-to-instable boundary follows a line through the turning points of the constant chemical potential curves. By analogy to the definition of the isothermal compressibility, $\beta_T = -\frac{1}{V}(\frac{\partial V}{\partial P})_T$, which measures the relative volume (*V*) change of matter by the applied pressure (*P*) at constant temperature (*T*), we define an iso-potential compressibility as $\beta_\mu = -\frac{1}{R}(\frac{\partial R}{\partial \alpha})_\mu$, which measures the relative change of pump spot (R) by the applied pump strength P/P$_{th}$ (proportional to the strength parameter $\alpha$) at constant chemical potential ($\mu$) in our system. Therefore, negative iso-potential compressibility defines the stability of the condensate, and the condensate behaves with positive compressibility on crossing from the stable to unstable region. In order to represent the strength parameter α in terms of the relative pump strength compared to the threshold value (P/P$_{th}$), we use the characteristic scales proposed in Ref. 14, assuming a pump power at twice threshold $\gamma_{eff} = \gamma - \kappa$ to be 0.13meV. Furthermore, the trap potential provided by disorder [14] or an opening hole surrounded by a thin metal (Ti/Au) film [9] is estimated to be 0.2 meV. Therefore, α can be mapped to P/P$_{th}$ by dividing a factor of 0.65 ( $\alpha = 2\gamma_{eff}/\hbar\omega = 0.65$ ).

The phase boundaries in Fig. 2 reveal not only the stable-to-instable transition but also a bifurcation of "soft" and "rigid" stable modes. The boundary can be characterized by the stability of the singly quantized vortex (excitation with the winding number *m* = $\pm$1). Just like the elementary excitations of the non-vortex states discussed in Section-2, we can find the solution of singly quantized vortex and then proceed the linear stability analysis. For the rigid mode, the singly quantized vortex is unstable. For the soft mode, however, the vortex is stable. Therefore, vortices can't survive in the rigid mode, but they can exist in the soft mode, forming vortex-antivortex pairs to exhibit the features of low viscosity in the superfluid. Therefore, the soft-rigid boundary represents the critical pumping size and strength for quasi-condensation which will become more clear with the concept of supercurrent.

## 4. The characteristics of the rigid and soft modes

We have solved the stationary solution by using the Madelung transformation. The supercurrent $V(\rho)$ can be defined as the gradient of the phase function $\nabla\varphi$, i.e., $V(\rho) = \nabla\varphi$. The rotationally symmetric distributions of steady-state densities, $A^2(\rho)$, the corresponding supercurrents, $V(\rho)$, and elementary excitations are shown in Fig. 3 for rigid ($R = L$) and soft modes ($R = 5\,L$). We found that not only the size and the density of soft mode are larger than that of the rigid mode but also the density profile of soft mode contains more obvious depletion in the radial position corresponding to a maximum flow velocity (at $r = 2\,L$ for soft mode and at $r = L$ for rigid mode). The non-equilibrium characters of the condensate create a nonzero supercurrent in the steady state so that the density distribution is affected by the supersurrent. Supercurrents are both zero outside and at the center of the condensate. From this boundary condition, there exist a maximal velocity somewhere in the middle of the condensate, where the density has to be depleted in order to preserve the chemical potential. So the radial position of the maximal supercurrent coincide with the position of density depletion. As for the direction of supercurrent is concerned, the density-dependent gain plays the important role. If the localized density experiences a net gain, the condensate tends to grow and flows outwards to sustain a constant density. However, when the localized density experiences a net loss, the condensate tends to be depleted and flows inwards to compensate the density. The soft mode is of the latter case, the supercurrent flows inwards, disturbing the density distribution more easily. The Goldstone flatness in the excitation spectrum also reveals that the excitations within the flatness can be excited without extra energy. Therefore, it's possible that the solution of vortex-antivortex pair exist in the soft mode as we found in the time-evolution of cGPE in Fig. 3(d) and our previous report [12]. On the contrary, for rigid mode, the density is not easily disturbed from the sense of zero-momentum excitation gap. When the system moves crossing the soft-rigid boundary by decreasing the pumping spot, the Goldstone flatness disappear that opens the excitation gap in the rigid mode accompanied with the instability of the singly quantized vortices. We therefore termed the condensate in the rigid mode as the local Bose condensation phase, and that in the soft mode as the quasi-condensate BKT phase [15, 16].

To understand the "quasi-condensate" BKT phase in the soft mode, the normalized correlation function is calculated using $g^{(1)}(r) = \langle \hat{\psi}^+(r')\hat{\psi}(r'-r) \rangle / n_0$, where $\hat{\psi}(r)$ is the quantum field operator and $n_0$ is the central condensate density. We can fit the spatial correlation function to a power-law relation $g^{(1)}(r) = r^{-\Lambda}$ and find the exponent as a function of the pumping spot. The exponent in our system has no jump and ranges from 0.4 to 0.5 as decreasing pumping spot from the soft-unstable boundary towards the rigid-soft boundary. On the contrary, the exponent in the trapped atomic system [5] changed from 0.46 to 0.29 on decreasing the temperature of the system through BKT transition. Another system on polariton condensates [17] discovered the exponent to be around 1.2 ~ 0.9 as a function of the pumping power. The exponent deviates from the equilibrium limit (1/4) that could come from the potential trap of the condensates, and the coherent feature becomes more apparent with decreasing pumping spot. In the soft-mode regime, the state shows short range correlation and the vortices would proliferate. The healing length $\xi = \hbar/\sqrt{2mUn}$, where n is the density of the condensate, goes linearly with the exponent $\Lambda$ since it represents the length scale over which the phase and density fluctuations can be suppressed by the interaction between condensed polaritons. As plotted in Fig. 4, the soft mode is characterized with Nambu-Goldstone mode that shows the long-wavelength excitations to destroy the coherence of the condensate. The flatness of the elementary excitations $\Delta k$ behaves the opposite trend of $\xi$ and so $\Lambda$. When the condensates are pushed into rigid-mode regime, the state shows long range correlations and the condensates start to exhibit the properties of superfluid. The analogy from the high temperature to low temperature transition is achieved in this work by varying the pumping spot size to decrease chemical potential of the polariton condensates.

**5. Physical interpretation of the localized BEC and BKT phase order**

The rigid mode is the regime of "localized BEC" in which the pump spot is comparable to the potential trap. The strict BEC implies the infinite and constant extension of coherent condensate in the direct space, however, in the finite-size system, the condensate as a whole is a single droplet that exhibits local Bose condensation. They should be characterized by a power-law decreasing correlation and the droplet size extends to the edge of pumping spot. The dispersion could show linear-type dispersion, and there is a ZMEE that reveals the rigidity of BEC-state. Moreover, vortices are not stable that further confirms the appearance of a homogeneous phase in direct space, i.e., superfluidity. For this small system (small pump spot), the first phenomenon that is encountered when the phase space density (by increasing pump strength) is increased is "conventional" BEC. For the soft mode, the condensates can be only considered as the "quasi-condensate" order. The spatial correlation length is short (smaller than the pump size), and the droplet size is correlated to the healing length of the wave-function. For a fixed pumping strength, the Goldstone flatness increases with the pumping spot size. We believe the flatness comes from the inward propagating supercurrent that is intrinsic to the pumped-dissipative system. Increasing pump size in the soft-mode regime, larger and more extended supercurrent move inside the condensate that contributes to more excitation components in the momentum space without extra energy. Therefore, the Goldstone flatness gets wider. For this larger system, the first relevant mechanism that occurs when increasing the phase space density is a BKT-like transition [23]. BKT transition comes first in the soft mode, the dispersion is flat which shows the Nambu-Goldstone behaviors .

## 7. Conclusion

In conclusion, we report the phase diagram of an exciton-polariton condensate in a harmonic confinement. On decreasing the pumping spot under a fixed pumping strength, the polaritons undergo a KT transition toward a quasi-condensate soft mode: the condensate exhibits Nambu-Goldstone excitations and short healing length. A further decrease of the pumping spot leads to the percolation process of the polariton condensate toward a local Bose condensation that exhibits superfluidity. The stability boundary is consistent with the sign change of an iso-potential compressibility. Moreover, within the stable regime, a crossover from the rigid to the soft mode can be distinguished by the stability of the singly quantized vortex as well as the shape of excitation spectrum. The rigid mode is the localized BEC phase with linear-type dispersion, the singly vortex is not stable, and elementary excitation has an excitation gap. The soft mode has flat dispersion with Goldstone flatness, while the inverse flatness is proportional to the healing length of the condensate. The singly quantized vortex can be stable. With these theoretical works, we confirmed the Berezinskii–Kosterlitz–Thouless-like phase order exists in a finite 2D trapped polariton system, and the criterion we searched in this paper can be used to further pursue vortex-antivortex pair dynamics associated with BKT transition.

We acknowledge the financial support from the National Science Council (NSC) of the Republic of China under Contract No. NSC99-2112-M-034-002-MY3 and NSC99-2112-M-009-009-MY3.

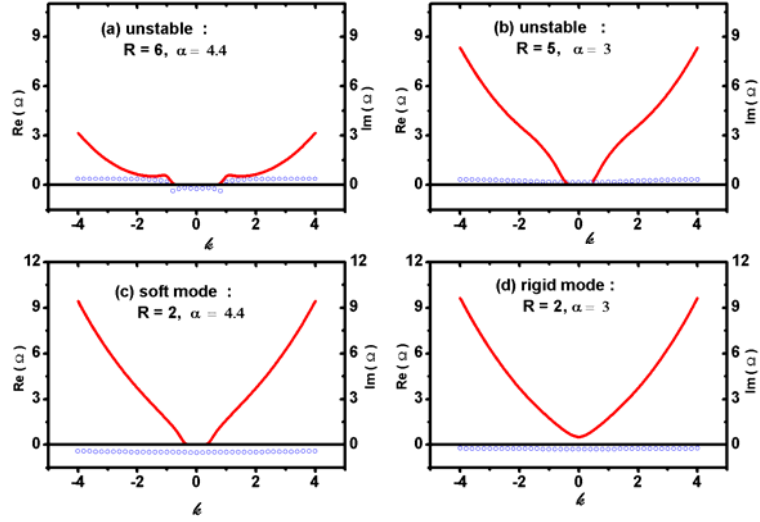

Fig. 1. Bogoliubov excitation spectra of stable and unstable modes. (a) Unstable mode for $R = 6$ and $\alpha = 4.4$, (b) unstable mode for $R = 5$ and $\alpha = 3$, (c) soft mode for $R = 2$ and $\alpha = 4.4$, and (d) rigid mode for $R = 2$ and $\alpha = 3$. Red lines indicate the real part and blue lines indicate the imaginary part of the eigenfrequency.

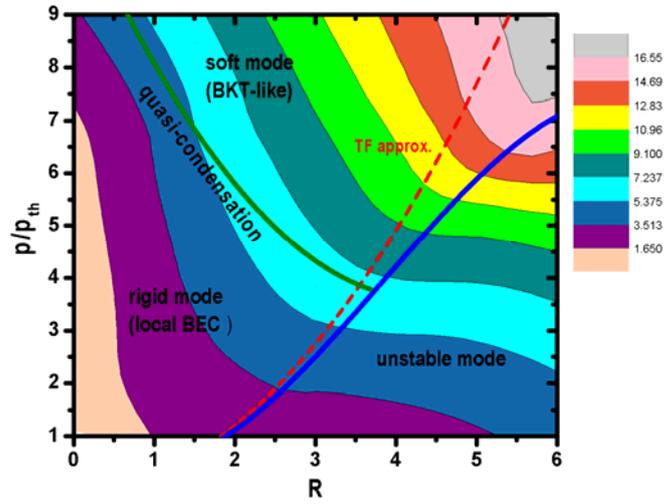

Fig. 2. Phase diagram for pump spot and pump strength with contour of chemical potential. Dashed red line is the Thomos-Fermi approximation boundary, and solid blue and green lines show the phase boundaries.

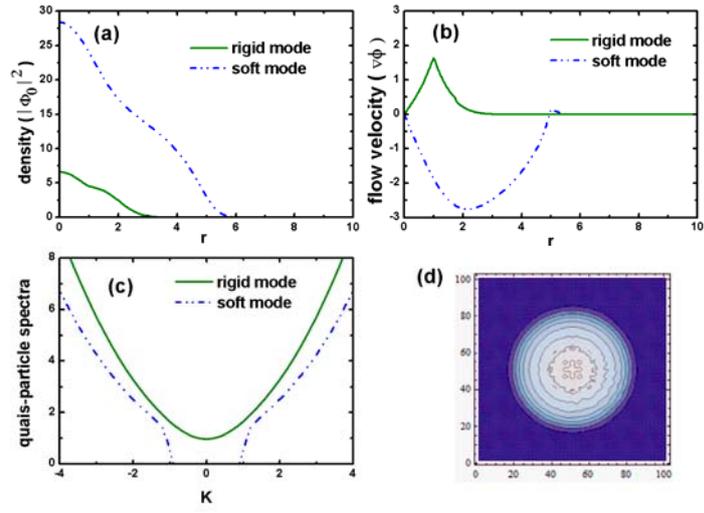

Fig. 3. The radial density, supercurrent distribution and elementary excitation of rigid and soft modes at $\alpha = 4.4$.

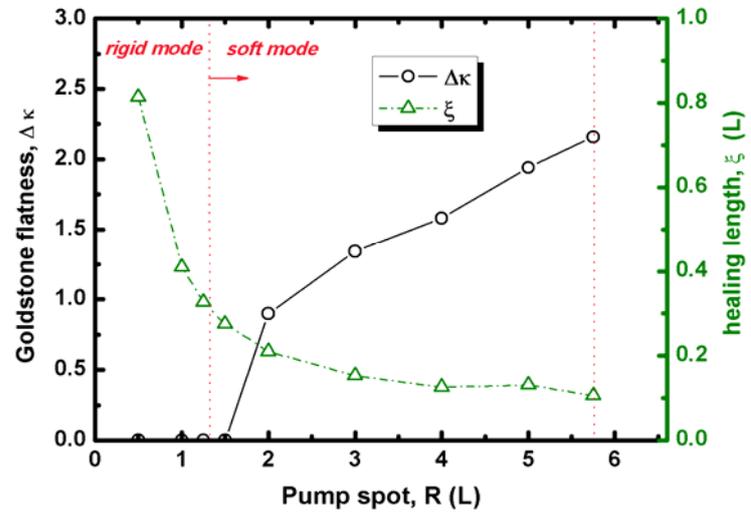

Fig. 4. Healing length $\xi$ and the Goldstone flatness $\Delta k$ as a function of pumping spots